\documentstyle[12pt,amssymb]{article}

\begin{document}

\tolerance=5000

\def\pp{{\, \mid \hskip -1.5mm =}}
\def\cL{{\cal L}}
\def\be{\begin{equation}}
\def\ee{\end{equation}}
\def\bea{\begin{eqnarray}}
\def\eea{\end{eqnarray}}
\def\beq{\begin{eqnarray}}
\def\eeq{\end{eqnarray}}
\def\tr{{\rm tr}\, }
\def\nn{\nonumber \\}
\def\e{{\rm e}}

\renewcommand{\title}[1]{\begin{center}\Large\bf #1\end{center}\rm\par\bigskip}
\renewcommand{\author}[1]{\begin{center}\Large #1\end{center}}
\newcommand{\address}[1]{\begin{center}\large #1\end{center}}
\def\dinfn{\small Dipartimento di Fisica, Universit\`a di Trento\\ 
                           and Istituto Nazionale di Fisica Nucleare,\\
                                   Gruppo Collegato di Trento, Italia \medskip}

\def\csic{\small Consejo Superior de Investigaciones Cient\'{\i}ficas\\
                        IEEC, Edifici Nexus 201, 
                            Gran Capit\`a 2-4, 08034 Barcelona, Spain,\\
                               and Departament ECM and IFAE, 
                                 Facultat de F\'{\i}sica,\\
                                      Universitat de Barcelona, 
                                 Diagonal 647, 08028 Barcelona, Spain \medskip}

\newcommand{\guido}{Guido Cognola\footnote{e-mail: 
\sl cognola@science.unitn.it\rm}}
\newcommand{\sergio}{Sergio Zerbini\footnote{e-mail: 
\sl zerbini@science.unitn.it\rm}}
\newcommand{\emilio}{Emilio Elizalde\footnote{e-mail: 
\sl elizalde@ieec.fcr.es, elizalde@math.mit.edu \rm\rm}}

\title{Multi-graviton theory from a discretized RS brane-world and the induced
cosmological constant}

\author{\guido$^a$, \emilio$^b$, 
Shin'ichi Nojiri\footnote{e-mail: \sl nojiri@nda.ac.jp, 
snojiri@yukawa.kyoto-u.ac.jp\rm}$^c$, \\  Sergei D. Odintsov\footnote{e-mail:
\sl odintsov@ieec.fcr.es Also at TSPU, Tomsk, Russia\rm}$^d$
and \sergio$^a$}

\address{$^a$  \small \dinfn \\  $^b$ \csic
\\  $^c$ Department of Applied Physics, 
National Defence Academy, \\
Hashirimizu Yokosuka 239-8686, JAPAN \medskip\\
$^d$ Institut d'Estudis Espacials de Catalunya (IEEC), \\
Edifici Nexus, Gran Capit\`a 2-4, 08034 Barcelona, Spain, \
 and \\
Instituci\`o Catalana de Recerca i Estudis 
Avan\c{c}ats (ICREA), Barcelona, Spain}

\bigskip \bigskip \bigskip 

\abstract{We propose a multi-graviton theory with non-nearest-neighbor 
couplings in the theory space. The resulting four-dimensional discrete 
mass spectrum
reflects the structure of a latticized extra dimension. For a plausible
mass spectrum motivated by the discretized Randall-Sundrum brane-world, 
the induced cosmological  constant turns out to be positive and may serve 
as a quite simple model for the dark energy of our accelerating universe.}

\newpage

\noindent{\bf 1. Introduction.} 
There is renewed interest in the study of multi-graviton
theories \cite{BDGH}. Generally speaking, the corresponding action 
can be viewed as a sum of
Pauli-Fierz terms while a consistent interaction of gravitons does not
occur there. In  a way,  multi-graviton theories resemble 
higher-dimensional gravities with a discretized dimension.
And such sort of discretized Kaluza-Klein theories are now under 
intense attention due to their primary importance in the realization of 
the dimensional deconstruction program \cite{akio,AH}. Moreover, 
multi-gravitons are also  related with discretized
brane-worlds \cite{DKP}.

In spite of the the absence of a consistent interaction among the gravitons,
one can think on possible couplings in the theory space. In particular, 
in a recent paper \cite{KS} a multi-graviton theory with nearest-neighbour 
couplings in the theory space has been proposed. As a result, a discrete 
mass spectrum appears. The theory seems to be equivalent to Kaluza-Klein 
gravity with a discretized dimension. 

In this letter, we propose a simple model of a multi-graviton theory with
non-nearest-neighbour couplings in the theory space. Depending on the choice
for such couplings, a the different graviton mass spectrum is induced.
For instance, our model may be seen as motivated by a discretized 
Randall-Sundrum (RS) brane-world \cite{RS1} or by complicated boundary condition on a latticized dimension. It opens number of possibilities for the choice 
of mass spectrum. For example motivated by discretized RS brane-world model 
the small positive cosmological constant is induced as shown by the explicit one-loop calculation.

\medskip 

\noindent{\bf 2.  A multi-graviton model.} 
We consider $N$ variables, $\phi_n$, which may be
identified with the fields on a lattice 
with $N$ sites. The difference operator $\Delta$ is defined by
\be
\label{KS1}
\Delta \phi_n \equiv \sum_{k=0}^{N-1}a_k \phi_{n+k}\ .
\ee
Here we assume $\phi_{n+N}=\phi_n$, which may be regarded as a 
periodic boundary  condition. In (\ref{KS1}), the $a_k$'s are $N$ constants. 
In order that $\Delta$ becomes a usual differentiation in a proper continuum 
limit, the following condition is imposed:
\be
\label{KS2}
\sum_{k=0}^{N-1}a_k = 0\ .
\ee
The eigenvectors for $\Delta$ are given by
\be
\label{KS3}
\phi_n^M={1 \over \sqrt{N}}\e^{i{2\pi nM \over N}}\ ,\quad M=0,1,2,3
\ee
and their corresponding eigenvalues, by
\be
\label{KS4}
\Delta \phi_n^M = im^M \phi_n^M\ ,\quad 
im^M=\sum_{n=0}^{N-1}a_n \e^{i{2\pi nM \over N}} \ .
\ee
Here $m^0=0$. We should also note that $\phi_n^M$ satisfies the following 
properties, which may be identified with the conditions of normalization 
and completeness, respectively:
\be
\label{KS5}
\sum_{n=0}^{N-1}\phi_n^{M*} \phi_n^{M'}=\delta^{MM'}\ ,\quad 
\sum_{M=0}^{N-1}\phi_n^{M*} \phi_{n'}^{M}=\delta_{nn'}\ .
\ee
One can solve $a_n$ with respect to $m^M$ by
\be
\label{KS6}
a_n = {i \over \sqrt{N}}\sum_{M=0}^{N-1} m^M \phi_n^{M*}
= {i \over N}\sum_{M=0}^{N-1} m^M \e^{-i{2\pi nM \over N}} \ .
\ee
Then by choosing $a_n$ properly, we may obtain arbitrary spectrum of $m^M$ with $m^0=0$. 
We should note if $a_n$ is real 
\be
\label{KS6aA}
m^{N-M}=-\left(m^M\right)^*\ .
\ee
In what follows, we will assume both that $a_n$ is real and that the 
condition (\ref{KS6aA}) is satisfied. 
For instance, if $N=2$, we find that $m^1$ is pure imaginary and that
\be
\label{KS6b}
a_0=-\Im m^1\ ,a_1=\Im m^1\ .
\ee
Here $\Im m^1$ is the imaginary part of $m^1$, and we denote the real part 
of $m$  by $\Re m$. For $N=3$, we have $m^2=-\left(m^1\right)^*$, and 
\bea
\label{KS6c}
&& a_0=-{2 \over 3}\Im m^1\ , \quad 
a_1 = {1 \over 3}\left(\Im m^1 + \sqrt{3} \Re m_1 \right)\ ,\nn
&& a_2={1 \over 3}\left(\Im m^1 - \sqrt{3} \Re m_1 \right)\ .
\eea
For $N=4$, we have $m^3=-\left(m^1\right)^*$ and find $m^2$ is pure imaginary. 
Then we obtain
\bea
\label{KS6d}
& a_0 = {1 \over 4}\left(2\Im m^1 - \Im m^2\right)\ ,\quad
& a_1 = {1 \over 4}\left(2\Re m^1 + \Im m^2\right)\ ,\nn
& a_2 = {1 \over 4}\left(-2\Im m^1 - \Im m^2\right)\ ,\quad
& a_3 = {1 \over 4}\left(-2\Re m^1 + \Im m^2\right)\ .
\eea
One may apply the above arguments to the multi-graviton theory
, extending the formalism in \cite{KS}. 

The Lagrangian of the massless spin-two field (graviton) $h_{\mu\nu}$ is 
given by
\be
\label{KS7}
{\cal L}_0=-{1 \over 2}\partial_\lambda h_{\mu\nu}\partial^\lambda h^{\mu\nu}
+ \partial_\lambda h^\lambda_{\ \mu}\partial_\nu h^{\mu\nu} 
 - \partial_\mu h_{\mu\nu}\partial_\nu h + {1 \over 2}\partial_\lambda h
\partial^\lambda h\ ,
\ee
where $h\equiv h^\mu_{\ \mu}$. The Lagrangian of the massive graviton with 
mass $m$ is  given by introducing St{\"u}ckerberg fields $A_\mu$ and 
$\varphi$ \cite{hamamoto}
\bea
\label{KS8}
{\cal L}_m &=& {\cal L}_0 - {m^2 \over 2}\left(h_{\mu\nu}h^{\mu\nu} - h^2\right) 
 - 2 \left(m A^\mu + \partial^\mu \varphi\right)
\left(\partial^\nu h_{\mu\nu} - \partial h\right) \nn
&& - {1 \over 2}\left(\partial_\mu A_\nu - \partial_\nu A_\mu\right)
\left(\partial^\mu A^\nu - \partial^\nu A^\mu\right) \ .
\eea
We consider $N$ copies of the graviton fields $h_{n\mu\nu}$ and 
St{\"u}ckerberg fields $A_{n\mu}$ and $\varphi_n$. 
By replacing 
\be
\label{KS8a}
h_{\mu\nu} \to h_{n\mu\nu}\ ,\quad A_\mu \to -i A_{n\mu}\ ,\quad \varphi \to \varphi_n\ ,
\quad m\to -i\Delta, 
\ee
and adding all terms in the action (sum on $n$), we hereby propose a theory
given by the following Lagrangian, which is a generalization of the one 
in \cite{KS}:
\bea
\label{KS9}
{\cal L}&=& \sum_{n=0}^{N-1}\left[ -{1 \over 2}\partial_\lambda h_{n\mu\nu}
\partial^\lambda h_n^{\mu\nu}
+ \partial_\lambda h^\lambda_{n\ \mu}\partial_\nu h_n^{\mu\nu} 
 - \partial_\mu h_{n\mu\nu}\partial_\nu h_n 
+ {1 \over 2}\partial_\lambda h_n\partial^\lambda h_n \right. \nn
&& + {1 \over 2}\left(\Delta h_{n\mu\nu}\Delta h_n^{\mu\nu} - \left(\Delta h_n\right)^2
\right) - 2 \left(\Delta^\dagger A_n^\mu + \partial^\mu \varphi_n\right)
\left(\partial^\nu h_{n\mu\nu} - \partial_\mu h_n\right) \nn
&& \left. + {1 \over 2}\left(\partial_\mu A_{n\nu} - \partial_\nu A_{n\mu}\right)
\left(\partial^\mu A_n^\nu - \partial^\nu A_n^\mu\right) \right]\ .
\eea
Here $\Delta$ is the difference operator, defined in (\ref{KS1}), which 
operates on the indices $n$, while $\Delta^\dagger$ is defined as
\be
\label{KS9a}
\Delta^\dagger \phi_n \equiv \sum_{k=0}^{N-1}a_k \phi_{n-k}\ ,
\ee
which satisfies the following equation:
\be
\label{KS9b}
\sum_{n=0}^{N-1}\phi^1_n \Delta \phi^2_n 
= \sum_{n=0}^{N-1}\left(\Delta^\dagger\phi^1_n\right) \phi^2_n \ .
\ee
The Lagrangian is invariant under  transformations with the local
parameters $\xi_n^\nu$ and $\zeta_n$:
\bea
\label{KS10}
h_{n\mu\nu} &\to& h_{n\mu\nu} + \partial_\mu \xi_{n\nu} + \partial_\nu \xi_{n\mu}\ ,\nn
A_{n\mu} &\to& A_{n\mu} + \Delta \xi_{n\mu} - \partial_\mu \zeta_n\ ,\nn
\varphi_n &\to& \varphi_n + \Delta^\dagger \zeta_n\ .
\eea
Since the spectrum of $\Delta$ is given by (\ref{KS4}), the Lagrangian 
describes a massless graviton (with mass $m^0=0$) and $N-1$ massive gravitons 
with masses $\left|m^M\right|$ ($M=1,2,\cdots, N-1$) in (\ref{KS4}). 
We should point out that the massive gravitons always appear in pairs, sharing 
a common mass.  As we have shown, the complex mass parameter  
$m^M$ ($M=1,2,\cdots, N-1$) can be arbitrary chosen, just by properly choosing 
the coefficients $a_k$ in (\ref{KS1}) to (\ref{KS6}). As discussed in 
\cite{KS}, the Lagrangian (\ref{KS9}) can be regarded as corresponding to 
a Kaluza-Klein theory where the extra dimension is a lattice. 

As an example, we consider the two-brane RS  model \cite{RS1}
(for a recent review, see \cite{maar1}). In that model, the masses $m^M$ of 
the Kaluza-Klein modes are given by the solutions of the following equation:
\bea
\label{KS11}
0&=&\left(y_2'(0) - {3 \over 2l}y_2(0)\right)
\left(j_2'\left(z_c\right) + {3 \over 2\left(l+z_c\right)}j_2\left(z_c\right)
\right)\nn
&&- \left(j_2'(0) - {3 \over 2l}j_2(0)\right)
\left(y_2'\left(z_c\right) + {3 \over 2\left(l+z_c\right)}y_2\left(z_c\right)
\right)\ .
\eea
Here $l$ is the length parameter of the five-dimensional AdS space and $z_c$ 
is given by 
the geodesic distance $\pi r_c$ between the two branes as 
\be
\label{KS12}
z_c=l\left(\e^{\pi r_c \over l} - 1 \right)\ .
\ee
The functions $j_2(z)$ and $y_2(z)$ are given in terms of Bessel functions as
\be
\label{KS13}
j_2(z)=\sqrt{z+l}\ J_2\left(m^M z + l\right)\ ,\quad
y_2(z)=\sqrt{z+l} \ Y_2\left(m^M z + l\right)\ .
\ee
Specifically, we can consider models given by  a first few solutions 
for $m^M$ of Eq.~(\ref{KS11}), in the different cases 
(\ref{KS6b}), (\ref{KS6c}), and (\ref{KS6d}). 
On the other hand, when $m^M$ is large, the Bessel functions behave as
\bea
\label{KS14}
\sqrt{z}\ J_2\left(m^M z\right) &\sim& \sqrt{2 \over \pi m^M}
\cos\left(m^M z - {5 \over 4}\pi\right)\ ,\nn 
\sqrt{z}\ Y_2\left(m^M z\right) &\sim& \sqrt{2 \over \pi m^M}
\sin\left(m^M z - {5 \over 4}\pi\right)\ .
\eea
Then Eq.~(\ref{KS11}) reduces to 
\be
\label{KS15}
\tan\left(m^M l - {5 \over 2}\pi\right)
\sim \tan\left(m^M \left(z_c + l\right) - {5 \over 2}\pi\right)\ ,
\ee
which has the solution 
\be
\label{KS16}
m^M={\pi K \over z_c}\ ,
\ee
$K$ being a positive integer. Motivated by (\ref{KS16}), we may consider a 
$N=2N'+1$ graviton model with graviton masses given by
\be
\label{RS17}
m^M=\left\{\begin{array}{ll} {\pi M \over z_c},\ & M=0,1,\cdots N', \nn
-{\pi(N-M)\over z_c},\ & M=N'+1, N'+2, \cdots N-1 =2N'. \nn
\end{array}\right.
\ee
 Eq.~(\ref{KS6}) gives
\begin{eqnarray}
a_0 &=& 0\ ,\nn  
a_n &=& - {2\pi\over (2N'+1)z_c}
\Im \left\{{ \left(1 - \e^{-i{2\pi N' n \over 2N'+1}}\right)\e^{-i{2\pi n 
\over 2N'+1}} \over 1 - \e^{-i{2\pi n \over 2N'+1}}}\right\} \nn
&=& -{(-1)^n\:2\pi\over N z_c}\:
\frac{\sin^2\left({\pi nN'\over N}\right)}
{\sin\left({\pi n\over N}\right)}\:, \quad (n\neq 0)\ .
\label{RS18}\end{eqnarray}
Then $N$ plays the role of a cutoff of the Kaluza-Klein modes. 

In previous models \cite{AH,KS} of deconstruction, mainly  nearest 
neighbour couplings  between the sites of the lattice had been considered. 
Then, on imposing a periodic boundary  condition, the lattice looks as a 
circle. Departing from this case, in the model considered here, we have  
introduced non-nearest-neighbour couplings among the sites. Then, a site 
links to a number of different ones in a rather complicated way. In this 
sense, the lattice in the present model is no more a simple  circle but it 
looks like, say, a mesh or a net. Let us assume that the sites on the lattice 
 correspond to points in a brane. If the codimension of 
the spacetime is one, the brane should be ordered, resembling
the sheets of a book. One brane can  only interact (directly) with the 
neighbouring two branes. However, if the spacetime is more complicated 
and/or the codimension is two or more, brane can (directly) interact with 
three or more branes, an interaction that will be perfectly described by 
our corresponding model. For example, a site on a tetrahedron connects
 with three neighbouring sites. In this way, the non-nearest couplings may 
adequately reflect the structure of the extra dimension. In this
respect our model is very general and opens a number of interesting 
possibilities.
 \medskip

\noindent{\bf 3. The induced cosmological constant.}
We will here compute the one-loop effective potential and
as a consequence the induced cosmological constant for the $N$
graviton model discussed in the previous section (for a recent review on
Casimir energy calculations from extra dimensions, see \cite{mil1}). 
We shall use zeta-function regularization \cite{zbs}, but instead, a different
 regularization scheme could work as well.  
First of all we compute the effective potential for 
a free scalar field with mass $M$ since this corresponds to the 
contribution of each degree of freedom to the one-loop effective potential
of our theory. 

In the zeta-function regularization scheme, 
the one-loop contribution to the effective potential is given by 
\bea
V^{(1)}_{eff}=-\frac1{2V}\zeta'(0)\:,
\label{v1a}\eea
$V$ being the volume of the manifold and $\zeta(s)$
the zeta function corresponding to the operator $L=-\nabla ^{2}+M^2$. 
In a flat manifold it can be easily computed by means of
\begin{eqnarray}
\zeta(s) &=& \frac{1}{\Gamma(s)}\int_0^\infty 
\,dt\:t^{s-1}\mbox{Tr}\:e^{-tL/\mu^2}=
\frac{V\mu^{2s}}{(4\pi)^{d/2}\Gamma(s)}\int_0^\infty 
\,dt\:t^{s-d/2-1}\:e^{-tM^2}
\nn 
&=&
\frac{V\:M^d}{(4\pi)^{d/2}}\:\frac{\Gamma(s-d/2)}{\Gamma(s)}
\left(\frac{M}{\mu}\right)^{-2s}\:,
\label{zAd}\end{eqnarray}
which is valid in arbitrary dimension $d$. The parameter $\mu$
has to be introduced for dimensional reasons and it will be fixed
by renormalization. 
We should notice that the last expression 
provides the analytic continuation of the zeta function 
to the whole complex $s-$plane. One can see that in even dimensions 
there are a finite number of simple poles at the integer points
$s=1,2,...,d/2$, while in odd dimensions there is an infinite
number of simple poles at all the half-integer points $s\leq d/2$.
In particular, in 4-dimensions the zeta function has only two simple
poles, and reads
\bea 
\zeta(s)=\frac{VM^4}{32\pi^2 (s-1)(s-2)}
\,\left(\frac{M}{\mu}\right)^{-2s}\:.
\label{z1}
\eea

By taking the derivative of Eq.~(\ref{zAd}) with respect to $s$, we obtain
\begin{eqnarray}
V^{(1)}_{eff}=\frac{(-1)^{d/2}M^d}{2(d/2)!(4\pi)^{d/2}}
\left[\ln\frac{M^2}{\mu^2}-\alpha(d)\right]\:,
\label{Veven}
\end{eqnarray}
which is valid in even dimensions, and 
\begin{eqnarray}
V^{(1)}_{eff}=-\frac{M^d\Gamma(-d/2)}{2(4\pi)^{d/2}}\:,
\end{eqnarray}
 valid in odd dimensions.  
Here $\alpha(d)=1+1/2+1/3+...+2/d$ is a rational number larger than 1. 

It has to be noted that, in the odd dimensional
case, the effective potential does not depend on $\mu$ and,
owing to the presence of the Euler's gamma function, it is alternatively
negative and positive. In particular, it is positive in five dimensions.
The situation drastically changes in even dimensions. In fact in such case
there is an explicit dependence on the scale parameter, and the effective
potential can be positive or negative, this depending  on the value
of $M/\mu$. 

In the multi-graviton model which we are considering in this paper, 
the effective potential can be obtained by adding up several 
contributions of the kind (\ref{Veven}) for $d=4$.
The masses are given by Eq.~(\ref{RS17}). In our model we have 
a massless graviton (which does not give contributions to the 
effective potential) and an even number of massive gravitons ( with 
the same mass). Then, in order to perform the computation, 
it is sufficient to consider only one-half of the whole mass spectrum,
that is
\be
M_p=\frac{\pi\:p}{z_c}\:,\qquad\qquad p=1,2,\:...\:,\frac{N-1}{2}\:.
\label{mp1a}\ee
In this way, we get
\begin{eqnarray}
V_{eff}&=&V_R(\mu)+10\sum_{p=1}^{\frac{N-1}2}\frac{M^4_p}{64\pi^2}
\left(\ln\frac{M_p^2}{\mu^2}-\frac{3}{2}\right)\nn
&=&V_R(\mu)+10\sum_{p=1}^{\frac{N-1}2}\frac{\pi^2p^4}{64z_c^4}
\left(\ln\frac{\pi^2p^2}{z_c^2\mu^2}-\frac{3}{2}\right)\nn
&=&V_R(\mu)+\frac{2^{-10}N(N^2-1)(3N^2-7)\pi^2}{3z_c^4}
\left(\ln\frac{\pi^2}{z_c^2\mu^2}-\frac{3}{2}\right)
\nn &&\qquad\qquad\qquad
+10\sum_{p=1}^{\frac{N-1}2}\frac{\pi^2p^4}{64z_c^4}\:\ln p^2\:,
\label{veff1a} 
\end{eqnarray}
where $V_R(\mu)$ is the renormalized vacuum energy and 
the factor of 10 comes from the fact that 
there are two gravitons with the same mass, each of them 
having 5 scalar degrees of freedom. 

The effective potential is a physical observable and for this reason 
it has not to depend on the choice of the arbitrary parameter scale $\mu$.
This means that it has to satisfy to some renormalization condition
(see e.g. \cite{cher1})
\beq 
\mu\:\frac{dV_{eff}}{d\mu}=0\:,
\eeq
from which it follows
\beq
V_R(\mu)=V_R(\mu_R)+
\frac{2^{-10}N(N^2-1)(3N^2-7)\pi^2}{3z_c^4}\ln\frac{\mu^2}{\mu_R^2}\:,
\label{vr1a}\eeq
$\mu_R$ being the renormalization point which can be fixed by the condition
$V_R(\mu_R)=0$. In this way one finally gets
\beq 
V_{eff}&=&\frac{2^{-10}N(N^2-1)(3N^2-7)\pi^2}{3z_c^4}
\left(\ln\frac{\pi^2}{z_c^2\mu_R^2}-\frac{3}{2}\right)
\nn &&\qquad\qquad\qquad
+10\sum_{p=1}^{\frac{N-1}2}\frac{\pi^2p^4}{64z_c^4}\:\ln p^2\:.
\label{V4}\eeq
It interesting to note that, whatever the value of $\mu_R$ is 
(in practice it is fixed by  some standard renormalization condition), 
due to the particular form of the mass spectrum, 
the expression (\ref{V4}) turns out to
be positive if the number of gravitons is sufficiently large.
This is a quite natural way to obtain a positive cosmological constant.
One can observe that for large values of $N$, 
the effective potential goes as $const\cdot N^5(\ln N)/z_c^4$.

Now let us consider (for comparison) the multigraviton model studied in 
Ref.~\cite{KS}. Also in that case there is a massless graviton, which can be 
omitted, and $N-1$ gravitons with masses
\begin{equation}
M_p  = 2m\sin\left(\frac{\pi p}{N}\right),
\qquad\qquad p=1,2,\:...,\:N-1\:.
\label{mp2}
\end{equation}
$m$ being a positive constant. From (\ref{z1}) and (\ref{Veven})
with $d=4$, it  directly follows that
\begin{equation}
\zeta(s)=\frac{m^4}{2\pi^2(s-1)(s-2)}
\left(\frac{2m}{\mu}\right)^{-2s}
\sum_{p=1}^{N-1}\sin^{4-2s}
\left(\frac{\pi p}{N}\right), 
\label{zp2}\end{equation}
and the corresponding effective potential, 
renormalized according to the procedure explained above, is
\begin{eqnarray}
V_{eff}&=&\frac{5m^4}{4\pi^2}\sum_{p=1}^{N-1}
\sin^4\left(\frac{\pi p}{N}\right)
\left[
\ln\frac{4m^2}{\mu^2_R}-\frac32
+\ln\sin^2\left(\frac{\pi p}{N}\right)\right]
\nn
&=&\frac{15N}{32\pi^2}\left(\ln\frac{4m^2}{\mu^2_R}-\frac32\right)
+\frac{5m^4}{2\pi^2}\sum_{p=1}^{N-1}
\sin^4\left(\frac{\pi p}{N}\right)
\ln\sin\left(\frac{\pi p}{N}\right)\:,
\label{zp01}
\end{eqnarray}
where the factor 5 reflects again the scalar degrees of freedom of 
massive gravitons. The sign of the effective potential for this second example 
(Eq.~(\ref{zp01})) can be positive or negative,  depending on the
value $m/\mu_R$, but, contrary to what happens in our model, it
is always negative, for the renormalization mass $\mu_R$
is large as compared with $m$ (see \cite{KS}).
The last expression has a  nice behavior with $N$ as $N\to\infty$,
namely $\sim\alpha N$, being $\alpha =-0.041180$.

We should now remark that these results correspond to our 
re-calculation of the model in \cite{KS}. In fact, we noticed that
a zeta function with a  strange behavior was obtained there: 
apparently the starting point was a standard 
representation for $\zeta(s)$, valid in 4 dimensions
for $\Re s>2$, but after analytic continuation 
a representation for $\zeta(s)$ valid only for $\Re s<2$ was obtained
(Eqs.~ (16) and (22) in \cite{KS}).
This was probably because of an unclear treatment of  
the massless graviton, which gives a divergent contribution in 
the original expression.
In particular, owing to the existence of this zero mode, it seems that
the step from Eqs.~(19) to (20) in \cite{KS} missed the fact
that the contribution of $p=0$ is special: all integer values of $\ell$ 
(and not only those which are multiple of $N$) contribute. This further 
yields a correction to Eqs.~(22) and (23) in \cite{KS}, which is exactly of
the same kind as the last term there but with $N=1$. Obviously,
this new contribution is infinite for $q=1$ (see the last term of
Eq. (23) in \cite{KS}). This is why our Eq.~(\ref{zp01})
differs from the corresponding one in \cite{KS}), what is rendered
clear when comparing the large $N$ behavior of both.



\medskip 

\noindent{\bf 4. Discussion.}
Let us now estimate the distance between the two branes in
our multi-graviton theory obtained from a discretized Randall-Sundrum model
in order to induce the observable cosmological constant.
We may write $V_{eff}$ as $V_{eff}=V_0 \left(N,z_c,\mu\right) z_c^{-4}$;
as the Hubble constant $H$ is $H\sim 10^{-33}$ eV in the present universe, 
the effective cosmological constant, $\Lambda_{eff}$, should be 
$\Lambda_{eff}\sim \left(10^{-33}{\rm eV}\right)^2$. 
Since the Planck scale is $10^{19}$GeV$\sim 10^{28}$eV, one gets  
$V_{eff}\sim {\Lambda_{eff} \over \kappa^2} 
\sim 10^{-33\times 2 + 28\times 2}\left({\rm eV}
\right)^4= 10^{-10}\left({\rm eV}\right)^4$. Then if 
$V_0 \left(N,z_c,\mu\right)$ is of order unity, we find that 
$z_c \sim 10^{2}$ to $10^{3}$ eV, and since the Planck length is 
$10^{-33}$ cm, the value of this $z_c$ corresponds to $10^{-3}$ to 
$10^{-2}$ cm.  In the (continuum) model  \cite{ADD}, if the number of  
extra dimensions is $n$, the length of the extra dimension will be 
$10^{{30 \over n} - 17}$ cm. Then, if $z_c$ corresponds to the length 
of the extra dimension, the value  obtained corresponds to the $n=2$ case. 
We  note that the value we have  obtained for $z_c$ is  within the 
present limits of the measurements.
 
As a consequence, the new multi-graviton theory we propose admits
quite an arbitrary mass spectrum. To be more specific, a model of this kind
may be induced by a discretized brane-world. The induced cosmological 
constant is then defined by quantum effects.
With a reasonable choice for the distance between the branes, the induced 
cosmological constant lies actually within the range of the admissible 
observational values. Thus, a
multi-graviton theory with non-nearest-neigbour couplings in the theory space
can provide a useful model for dark energy 
in the current accelerating universe. The whole approach may open a
 new window in the realization of the deconstruction program for QED/QCD 
like theories. Indeed, by similar adjustments of the mass spectrum one 
is able to obtain  quite noticeable 
changes in the four-dimensional scalar effective potential. 
And such a dependence of the effective potential on the choice of 
mass spectrum allows us directly to put 
bounds to reasonable latticized boundary conditions from 
phenomenologically accepted estimations of the four-dimensional scalar 
potential.

\medskip 

\noindent{\bf Acknowledgments}
This investigation has been supported by the Program
INFN(Italy)-CICYT(Spain), by DGICYT (Spain), project BFM2000-0810 
 and by the Ministry of
Education, Science, Sports and Culture of Japan under the grant n.13135208
(S.N.).

\end{document}